\documentclass[aip,pop,sd,reprint,10pt,amsmath,amssymb,graphicx]{revtex4-1}

\usepackage{bm}
\usepackage{dcolumn}
\usepackage{graphicx}
\usepackage{color}
\usepackage{caption}
\usepackage{subcaption}
\usepackage{amsmath}
\usepackage{mathtools}
\usepackage{amssymb}

\newcommand{\nc}{\newcommand}
\nc{\rcite}[1]{Ref.~\onlinecite{#1}}
\nc{\rcites}[1]{Refs.~\onlinecite{#1}}
\nc{\eqeqref}[1]{Eq.~\eqref{eq:#1}}
\nc{\eqseqref}[2]{Eqs.~\eqref{eq:#1}-\eqref{eq:#2} }
\nc{\secref}[1]{Sec.~\ref{sec:#1}}
\nc{\secsref}[2]{Sec.~\ref{sec:#1}-Sec.~\ref{sec:#2}}
\nc{\ssecref}[1]{Sec.~\ref{ssec:#1}}
\nc{\ssecsref}[2]{Sec.~\ref{ssec:#1}-Sec.~\ref{ssec:#2}}

\begin{document}
\title{A Hamiltonian Five-Field Gyrofluid Model}
\author{I. Keramidas Charidakos, F. L. Waelbroeck and P. J. Morrison}
\affiliation{Institute for Fusion Studies and Department of Physics, The University of Texas at Austin, Austin, TX 78712, USA}

\begin{abstract}
A Lie-Poisson bracket is presented for a five-field gyrofluid model, thereby showing the model to be Hamiltonian. The model includes the effects of magnetic field curvature and describes the evolution of the electron and ion gyro-center densities, the parallel component of the ion and electron velocities, and the ion temperature. The quasineutrality property and Ampere's law determine respectively the electrostatic potential and magnetic flux. The Casimir invariants are presented, and shown to be associated to five Lagrangian invariants advected by distinct velocity fields. A linear, local study of the model is conducted both with and without Landau and diamagnetic resonant damping terms. Stability criteria and dispersion relations for the electrostatic and the electromagnetic cases are derived and compared with their analogs for fluid and kinetic models.
\end{abstract}
\maketitle

\section{INTRODUCTION}

Reduced electromagnetic fluid models constitute versatile tools for the study of multi-scale phenomena including in particular the interaction of turbulence with magnetohydrodynamic perturbations exhibiting meso-scale structures.\cite{FLW_IISS} Examples include magnetic islands,\cite{ishizawa2013magnetic,hillhariotta} edge localized modes,\cite{xu2013gyro,xi2014phase} resonant magnetic perturbations,\cite{militello2009error,ComiXTaylr} as well as fishbone\cite{odblom2002nonlinear} and Alfv\'{e}n modes.\cite{Spong94,tobias2011fast} Irrespective of the phenomenon that a particular fluid model aims to describe, the underlying system of charged particles interacting with electromagnetic fields is a Hamiltonian system and in addition to energy and other invariants related to symmetry properties, it may possess approximate Poincar\'e or adiabatic invariants such as wave actions. It is highly desirable that a model giving a reduced description should retain this important property in the ideal limit. By ideal limit, we mean the limit of the model when all dissipative terms, such as collisions, Landau damping and dissipative anomalous transport terms are neglected. Thus, Hamiltonian systems conserve energy for closed boundary conditions and the Hamiltonian formulation is useful for investigating the local properties of the dynamics that are independent of the drive.

Casting a system into its Hamiltonian form\cite{morrison1982hamiltonian,morrison1998hamiltonian} confers several practical advantages. One of the most important is the existence of families of invariants, called Casimir invariants, which are found in noncanonical Hamiltonian systems due to the degeneracy of the cosymplectic matrix. The functional that results from the addition of the Casimirs to the Hamiltonian has non-trivial equilibrium states as stationary points. In the absence of a Poisson bracket, by contrast, the existence of non-trivial equilibrium states is not guaranteed. For example, \rcite{waelbroeck2004hamiltonian} presents an example of a seemingly reasonable fluid model that lacks physical equilibria with  closed streamlines because the equilibrium equations imply that some fields are multiple-valued on closed streamlines. We can also take advantage of the Hamiltonian formulation to construct ``energy principles" for the investigation of the stability of such non-trivial equilibrium states by examining the second variation of the aforementioned functional.\cite{andreussi2013hamiltonian,morrison2013energy,tronci2015energy} Another advantage is that imposing constraints on a system is straightforward in the Hamiltonian formalism.\cite{chandre2012hamiltonian} Lastly, the Hamiltonian formalism can be used to facilitate the calculation of the statistical average of the zonal flow growth rate.\cite{Krommes2004}

Among the several classes of fluid models, of particular importance are the ones that retain the effects of finite ion temperature, principally for describing instabilities with growth rates comparable to the ion diamagnetic frequency or modes with perpendicular wavelengths of  the order of the ion Larmor radius. Whereas ``cold ion" models have been shown to possess noncanonical Hamiltonian formulations,\cite{morrison1984hamiltonian,tassi2008hamiltonian} the task of formulating such ``hot-ion" models that satisfy the Hamiltonian property has proven difficult. For example, efforts to identify the Hamiltonian structure of the four-field model of \rcite{hazeltine1985four}were unsuccessful, even though it conserves energy.\cite{,hazeltine1985fourE}  The main difficulty with such models lies in the nonlocality of the ion dynamics caused by Larmor gyration. One way to approximate nonlocal terms is by a Taylor-series, using $k_\perp \rho_i$ as a small parameter. An example of such a so-called FLR model  was given  in \rcite{hazeltine1987hamiltonian},  where a Hamiltonian four-field model is constructed, using the ``gyromap" technique to introduce finite ion temperature into the cold ion limit of \rcite{hazeltine1985four}. Unfortunately, we are unaware of any numerical implementation of this model, possibly because it requires high-order derivatives and, consequently, additional boundary conditions. 

An alternative approach for constructing fluid models with a finite ion temperature is to truncate the moment hierarchy of the gyrokinetic equation.\cite{dorland1993gyrofluid,brizard1992nonlinear,snyder2001landau,scott2005free,scott2010derivation} This leads to the use of nonlocal averaging operators that account for the full range of perpendicular wavelengths. The resulting models are called gyrofluid models. Surprisingly, gyrofluid models are more readily amenable to Hamiltonian formulations than FLR models. Examples of Hamiltonian electromagnetic gyrofluid models are given in \rcite{waelbroeck2009hamiltonian} for an incompressible (three fields) and  \rcite{waelbroeck2012compressible} for a compressible (four fields)  model. The four-field gyrofluid model advances the first two moments of the distribution function for each species, or the ion and electron densities and parallel momenta. Zacharias {\em et al.} have shown that simulations of magnetic reconnection using this model are in good agreement with gyrokinetic simulations,\cite{zacomi14} and Comisso {\em et al.} have used it to bring to light the acceleration of magnetic reconnection by nonlocal gyrofluid effects.\cite{gyro-acceln} Grasso {\em et al.}, by contrast, have used it to examine the stabilizing effects of ion diamagnetic drifts on the growth and saturation of tearing modes in inhomogeneous plasma.\cite{Gra12}

In the present paper, we propose a Hamiltonian five-field electromagnetic gyrofluid model that is an extension of the model presented in \rcite{waelbroeck2012compressible}. The new model, like its predecessor, is a truncation of a more complete one proposed by Snyder and Hammett, which advances six moments for the ions and two moments for the electron dynamics.\cite{snyder2001landau} We note that Scott\cite{scott2005free,scott2010derivation} has shown that achieving energy conservation requires modifying several of the terms in \rcite{snyder2001landau} involving higher order moments. We will likewise show that constructing a {\em Hamiltonian} model requires modifying the terms involving the higher order moments in our model.
 The new model extends that in \rcite{waelbroeck2012compressible} by the addition of the evolution of the ion temperature. As in the previous model, ion compressibility effects and field curvature are also included, allowing it to describe ITG, KBM, drift waves and tearing modes. To demonstrate the properties of the model, we present a linear, local study of electrostatic slab ITG and toroidal electromagnetic ITG modes.

The paper is organized as follows: In Section \ref{sec:model equations} we give the normalizations of our variables and present the ideal limit of the dynamical model. In Section \ref{sec:Hamiltonian} we give the Hamiltonian formulation of the model equations by providing a conserved energy that serves as the Hamiltonian and a Lie-Poisson bracket that satisfies the Jacobi identity. In Section \ref{sec:Casimirs} we calculate the Casimir invariants of our system and from them, in Section \ref{sec:normal fields} we construct five ``normal fields" which are field variables in which the dynamical equations and the bracket take a very simple form. Lastly, in Section \ref{sec:linear study} we perform a local, linear study of the model with particular emphasis on the study of the ITG and KBM modes. We present stability criteria for both the ideal model and a model with linear dissipation terms representing the effects of parallel Landau damping and the drift resonance. We investigate several well known stabilizing factors of the instability to show qualitative agreement with kinetic models.

\section{IDEAL MODEL}\label{sec:model equations}

We first present the ideal portion of our model  by omitting  collisional diffusion and wave-particle interaction terms, which  will  be examined  in Sec.~\ref{sec:linear study}. 

We are interested in a model that describes the destabilization of the drift wave excited by the ion temperature gradient. Due to the acoustic nature of the instability, we cannot neglect ion motion along the field lines;  therefore, we keep ion compressibility effects. Also, because we want to investigate toroidal plasma with finite $\beta$, we include electromagnetic effects. Lastly, to represent the influence of toroidicity, we allow for magnetic curvature. 
We consider the evolution of the the magnetic flux $\psi$, of a magnetic field $\bf{B} = \hat{\bf{z}} + \nabla\psi\times\hat{\bf{z}}$, the ion density $n_i$, the parallel velocity of the ion \text{\it{guiding centers}} $u_i = \hat{\bf{z}}\cdot\bf{v}_i$, the electron density $n_e$ and parallel velocity $u_e = \hat{\bf{z}}\cdot\bf{v}_e$, the electrostatic potential $\phi$ and the parallel ion temperature $T_\parallel$. We normalize these quantities in the following way:
\begin{eqnarray}
\lefteqn{
(n_i,n_e,\psi,\phi,u_i,u_e,T_{\parallel}) = }\nonumber\hspace{9mm} \\ 
& &\frac{L_n}{\rho_i}\left(\frac{\hat{n}_i}{n_o},\frac{\hat{n}_e}{n_o},\frac{\hat{\psi}}{\rho_i B_o},\frac{e\hat{\phi}}{\tau T_i},\frac{\hat{u}_i}{v_{ti}},\frac{\hat{u}_e}{v_{ti}},\frac{\hat{T}_{\parallel}}{T_i}\right)\,,
\label{normalizations}
\end{eqnarray}
where the carets denote the dimensional variables. Here $n_o$, $B_o$ and $T_e$ are the background density, magnetic field and electron temperature, $\rho_i = v_{ti}/\omega_{ci}$ is the ion Larmor radius, where $v_{ti} = (T_i/m_i)^{\frac{1}{2}}$ is the ion thermal speed, $\omega_{ci} = e B_o/m_i$ is the ion cyclotron frequency, $L_n = n_o/|\nabla n|$ is the density scale-length and $\tau = T_e/T_i$ is the ratio of the species temperatures.
We also normalize the independent variables according to:
\begin{equation}
(t,k_\parallel,k_\perp) = \left(\frac{\hat{t}v_{ti}}{L_n},\hat{k}_\parallel L_n,\hat{k}_\perp\rho_i\right)\,.
\label{ind_var_normal}
\end{equation}

With these normalizations, our evolution equations are as follows. The equations that describe the ideal evolution of ion quantities are
\begin{align}
&\frac{dn_i}{dt} = -\nabla_\parallel u_i - 2 u_d \frac{\partial}{\partial y} (n_i + \Phi + T_\parallel)\,,  \label{n_i}\\
&\frac{d(\Psi + u_i)}{dt} = -\nabla_\parallel T_\parallel  - \nabla_\parallel n_i -4 u_d \frac{\partial u_i}{\partial y}\,,   \label{M_i}\\
&\frac{dT_\parallel}{dt} =  -(\gamma-1)\nabla_\parallel u_i - 2u_d\frac{\partial}{\partial y} (n_i+\Phi+  T_\parallel)\,,
\label{T}
\end{align}
whereas the equations describing the evolution of electron quantities are
\begin{align}
&\frac{dn_e}{dt} = -\nabla_\parallel u_e +2u_d \frac{\partial}{\partial y}(n_e-\phi)\,, \label{n_e}\\
&\frac{ d(\psi - \mu u_e)}{dt} = \frac{1}{\tau}\nabla_\parallel n_e + 2\mu u_d\frac{\partial u_e}{\partial y}\,.
 \label{M_e}
\end{align}
In Eqs.\ \eqref{n_i}--\eqref{M_e},   $df/dt= \partial f/\partial t + [\Phi,f]$ and   $\nabla_\parallel f = \partial f/\partial z - [\Psi,f]$, with 
 $[\cdot,\cdot]$ denoting  the canonical Poisson bracket, so that $[f,g] = \hat{\bf{z}}\cdot(\nabla f\times\nabla g)$.  Also,  $\gamma$ is the adiabatic index, $u_d = L_n/R$ is the normalized curvature drift velocity, $R$ is the radius of curvature of the magnetic field and $\Phi = \Gamma_o^{1/2}\phi$, $\Psi = \Gamma_o^{1/2}\psi$ are the gyro-averaged $\phi$ and $\psi$. The symbol $\Gamma_o^{1/2}$ refers to the gyroaveraging operator introduced  in \rcite{dorland1993gyrofluid} and is defined by 
\begin{equation}
\Gamma_o^{1/2}\xi = \text{exp}\left(\frac{1}{2}\nabla^2_\perp\right)I^{1/2}_o\left(-\nabla^2_\perp\right)\xi \,,
\label{Gamma_def}
\end{equation}
where $I_o$ is a modified Bessel function of the first kind and the result of Eq.\eqref{Gamma_def} should be interpreted in terms of its series expansion. At this point, we note that only the ion guiding centers respond to the gyroaveraged value of the electromagnetic field. Therefore, we are required to use the gyroaveraged value of the electrostatic potential in the $\bf{E}\times \bf{B}$ drift advecting the ions whereas, electrons are advected only by the local value of their $\bf{E}\times \bf{B}$ drift since we neglect the electron Larmor radius. 

Equations \eqref{n_i}-\eqref{M_e} are closed by the parallel component of Amp\'{e}re's law
\begin{equation}
\frac{2}{\tau\beta_e}\nabla^2_{\perp}\psi = -j = -\Gamma_o^{1/2}u_i + u_e,\label{Ampere}
\end{equation}
with $j = \hat{\bf{z}}\cdot \bf{J}$ being the z-component of the current density, and by the quasineutrality condition
\begin{equation}
n_e = \Gamma_o^{1/2}n_i + \left(\Gamma_o -1\right)\phi,\label{quasineutrality}
\end{equation} 
with $\Gamma_o = \left(\Gamma^{1/2}_o\right)^2$. Here, $\Gamma_o^{1/2}n_i$ is the gyrophase-independent part of the real space ion particle density and the $(\Gamma_o -1)\phi$ term comes from the gyrophase-dependent part of the distribution function. It represents the ion polarization density due to the variation of the electric field around a gyro-orbit. We leave $\beta_e$ unrestricted so that we can describe both ``inertial" ($\beta_e\ll \mu$) and ``kinetic" ($\beta_e\gg \mu$) Alfv\'{e}n waves.
Since our only temperature equation involves the parallel temperature, from now on we will drop the subscript from $T_{\parallel}$.

It is interesting to compare the model presented in equations \eqref{n_i}-\eqref{M_e}
to one obtained from the models of \rcites{snyder2001landau,scott2005free,scott2010derivation}
by discarding all the terms involving high-order moments and associated terms. By ``associated'' terms, we mean for example that discarding $T_{\bot}$ requires that one also discard terms involving the gyroaveraging operator $J_1$, since the latter terms result from the effects on gyroaveraged quantities of the variations in the perpendicular temperature. The link between $T_\bot$ and $J_1$ is reflected in the fact that for energy conservation, $J_1$ terms must appear together with $T_\bot$, as noted in \rcites{scott2005free,scott2010derivation}. The omission of the terms containing $J_1$ means, in effect, that we neglect $\nabla J_0$. 
Compared to such a truncated model, the Hamiltonian model in Eqs.~\eqref{n_i}-\eqref{M_e} lacks any trapped particle effects (terms proportional to $\nabla_\| B$ in \rcites{snyder2001landau,scott2005free,scott2010derivation}) and has a less accurate treatment of FLR terms (due to the omission of the $J_1$ terms). The two models also differ in the coefficients of the various curvature terms. In the continuity equation, for example, the argument of the curvature operator in the truncated version of the model of \rcites{snyder2001landau,scott2005free,scott2010derivation} is $\Phi+p_\|/2$, while that in our model is $\Phi+p_\|$. This difference is necessary in order for the five-field model to conserve energy. In fact, we note that the curvature terms in Eqs.~(\ref{n_i}), (\ref{T}) and (\ref{n_e}) are the same as the ones found in the corresponding equations of the FLR fluid model of \rcite{zeiler1997nonlinear}, which evolves three ion moments, as we do, and conserves energy. Lastly, we note that the factor of four in front of the curvature term in the momentum equation, Eq.~(\ref{M_i}), does match the corresponding term in \rcites{snyder2001landau,scott2005free,scott2010derivation} despite the fact that for the four-field model of \rcite{waelbroeck2012compressible}, satisfying the Jacobi identity required halving this factor. The conclusion of these observations is that constructing Hamiltonian models requires modifying the truncated moment expansions, but that the correct terms are recovered as on increases the order of the model.    

\section{THE HAMILTONIAN FORM}\label{sec:Hamiltonian}

The system described in Sec.~\ref{sec:model equations}  conserves the following energy:
\begin{eqnarray}
\lefteqn{H = \frac{1}{2}\int_{\mathcal{D}} d^2x \left( \frac{n^2_e}{\tau} + n^2_i +\frac{1}{\gamma-1} T^2 + \mu u^2_e + u^2_i  \right.}\nonumber\hspace{25mm}\\
& \left.+\frac{2}{\tau\beta_e}|\nabla\psi|^2 + \Phi n_i - \phi n_e\right)\,,
\label{Hamiltonian}
\end{eqnarray}
where $\mathcal{D}$ denotes the spatial domain of interest and the boundary conditions are such that surface terms vanish. The successive terms of the functional of Eq.\eqref{Hamiltonian} represent, respectively, the electron and (two terms) ion thermal energies, the parallel component of the electron and ion kinetic energies, the magnetic energy and the electrostatic energies of ions and electrons. 
Taking the energy functional as the Hamiltonian of our 5-field model, we can write the set of equations in a noncanonical\cite{morrison1998hamiltonian} Hamiltonian form
\begin{equation}
\frac{\partial \xi^i}{\partial t} = \{\xi^i,H\}, \quad i = 1,\ldots,5,\label{Ham_form}
\end{equation}
with $\xi^i$ being the field variables and $\{\cdot,\cdot\}$  being  a non-canonical Poisson bracket. We employ the dynamical variables $n_i, M_i, n_e, M_e, T$, where $M_i = \Gamma_o^{1/2} \psi +u_i$ is the canonical ion momentum and $M_e = \psi -\mu u_e$, the electron one. Additionally, we define $\tilde{n}_i = n_i-2u_d x$, $\tilde{n}_e = n_e-2u_d x$ and $\tilde{T} = T-2u_d x$ for convenience. In these variables, the bracket given by
\begin{align}
\{F,G\} =& \int\! d^3x \Big( -\tilde{n}_i([F_{n_i},G_{n_i}]+[F_{M_i},G_{M_i}] 
\notag\\&+[F_{T}, G_{T}])  - M_i([F_{M_i},G_{n_i}] + [F_{n_i},G_{M_i}] 
\notag\\& 
+([F_{T},G_{M_i}]+[F_{M_i},G_{T}])) \notag\\&- \tilde{T} ([F_{n_i},G_{T}]+[F_{T},G_{n_i}]+[F_{M_i},G_{M_i}]) \notag\\&+ \tilde{n}_e ([F_{n_e},G_{n_e}]+\mu[F_{M_e},G_{M_e}]) \notag\\&+ M_e([F_{M_e},G_{n_e}] + [F_{n_e},G_{M_e}])\notag\\&
-(F_{M_i}\partial_z G_{n_i} - G_{M_i}\partial_z F_{n_i})\notag\\&
-(F_{T} \partial_z G_{M_i} - G_{T} \partial_z F_{M_i})\notag\\&
+(F_{M_e}\partial_z G_{n_e} - G_{M_e}\partial_z F_{n_e})\Big)
\label{bracket} 
\end{align}
satisfies the formulation of Eq.~\eqref{Ham_form} for the Eqs.~\eqref{n_i}--\eqref{M_e}, is bilinear, antisymmetric and satisfies the Jacobi identity. In the above bracket, we have taken $\gamma =2$ because this  is the only value of the adiabatic index that allows the bracket to satisfy the Jacobi identity, as shown by a 
direct proof of the Jacobi identity  using the techinques of \rcite{morrison1982hamiltonian}.  The Jacobi for this case will become evident in Sec.~\ref{sec:normal fields}.

\section{CASIMIR INVARIANTS}\label{sec:Casimirs}

One of the most important properties of noncanonical Hamiltonian systems is the existence of Casimir invariants, that is, constants  of  motion for any choice of Hamiltonian. A Casimir invariant $C$ thus needs to satisfy the relation $\{F,C\} = 0$ for any field $F$. Here, we will set $\partial_z = 0$. The generalization is straightforward. 

Assuming a  Casimir functional $C(n_i,M_i,T,n_e,M_e)$ and applying the condition $\{\xi_j,C\} = 0$ with $\xi_1 = n_i$, $\xi_2 = M_i$, $\xi_3 = T$, $\xi_4 = n_e$, $\xi_5 = M_e$ gives following:
\begin{align}
&[n_i-2u_d x, C_{n_i}] + [M_i, C_{M_i}] + [T-2u_d x,C_{T}] = 0 \label{Cas_sys_1}\\
&[n_i-2u_d x, C_{M_i}] + [M_i, C_{n_i}]  \nonumber\\
&\hspace{2cm} + [T-2u_d x, C_{M_i}]  + [M_i, C_{T}] = 0 \label{Cas_sys_2}\\
&[n_i-2u_d x, C_{T}] + [T-2u_d x, C_{n_i}] + [M_i, C_{M_i}] = 0 \label{Cas_sys_3}\\
&[n_e-2u_d x, C_{n_e}] + [M_e, C_{M_e}] = 0\label{Cas_sys_4}\\
&\mu[n_e-2u_dx, C_{M_e}] + [M_e, C_{n_e}] = 0\,.
\label{Cas_sys_5}
\end{align}

For the rest of this section, we employ the previously defined variables $\tilde{n}_i$, $\tilde{n}_e$, $\tilde{T}$. In addition, we observe that $F_{\tilde{\xi}} = F_{\xi}$. From  \eqref{Cas_sys_1} and  \eqref{Cas_sys_5} we retrieve no information since they are automatically satisfied for any choice of $C$.
However, from  \eqref{Cas_sys_2} we get 
\begin{align}
&[\tilde{n}_i,M_i](C_{M_i M_i} - C_{n_i n_i} - C_{T n_i}) \notag\\+ &[M_i,\tilde{T}](C_{n_i T} - C_{M_i M_i} + C_{T T})\notag\\ + &[\tilde{n}_i,\tilde{T}](C_{M_i T} - C_{M_i,n_i}) = 0\,,
 \label{Cas_1}
\end{align}
from  \eqref{Cas_sys_3} we get
\begin{align}
&[\tilde{n}_i,\tilde{T}](C_{T T} - C_{n_i n_i})\notag\\ + &[\tilde{T},M_i](C_{n_i,M_i} - C_{M_i T})\notag\\ + &[\tilde{n}_i,M_i](C_{T M_i} - C_{M_i n_i}) = 0\,,
\label{Cas_2}
\end{align}
and from \eqref{Cas_sys_4} we get
\begin{equation}
[\tilde{n}_e,M_e](\mu C_{M_e M_e} - C_{n_e n_e}) = 0\,.
\label{Cas_3}
\end{equation}
Accordingly, we have the following set of equations:
\begin{align}
C_{M_i M_i} - C_{n_i n_i} - C_{T n_i} &= 0\label{Cas_eq1}\\
C_{M_i M_i} - C_{n_i T} - C_{T T} &= 0\label{Cas_eq2}\\
C_{T M_i} - C_{M_i n_i} &= 0\label{Cas_eq3}\\
C_{T T} - C_{n_i n_i} &=0\label{Cas_eq4}\\
\mu C_{M_e M_e} - C_{n_e n_e} &= 0\,,
\label{Cas_eq5}
\end{align}
which must be satisfied by any Casimir invariant.

We start from Eq.\eqref{Cas_eq3} and integrate it w.r.t $M_i$ to find $C_{n_i} = C_{T} + f(\tilde{n}_i,\tilde{T})$. By using the method of characteristics on this result, we infer that the solution has the form $C =\left< g(\tilde{T}+\tilde{n}_i,M_i) + f(\tilde{n}_i,\tilde{T})\right>$, where the $\left<\right>$ symbol implies an integral over the volume of interest. Subsequently, we substitute this form of the Casimir into \eqref{Cas_eq4} to obtain the wave equation $\partial^2_{n_i}(f+g) - \partial^2_{T}(f+g)=0$ and by application of the method of characteristics, we recover the other characteristic direction, $C = \left<g(\tilde{T}+\tilde{n}_i,M_i) + f(\tilde{n}_i - \tilde{T})\right>$. Finally, employing  \eqref{Cas_eq1} we arrive at the wave equation $\partial^2_{M_i} g - 2\partial^2_{n_i + T} g =0$. Invoking the method of characteristics once more, we derive the following general form for the Casimir invariants corresponding to the ion piece of the bracket: 
\begin{equation}
C_i = \int \text{$d^2$x} \quad g_\pm(\tilde{T}+\tilde{n}_i \pm \sqrt{2}M_i)  + f(\tilde{n}_i - \tilde{T})\,.
\end{equation}

For the Casimir invariants that correspond to the electron part of the bracket, we need only  solve  \eqref{Cas_eq5} to obtain 
\begin{equation}
C_e = \int \text{$d^2$x} \quad h_\pm(M_e \pm \sqrt{\mu}\tilde{n}_e) \,.
\end{equation}

Thus, a  general family of  Casimir invariants is given by  
\begin{align}
C(n_i,M_i,T,n_e,M_e) =& \int \text{$d^2$x} \quad g_\pm(\tilde{T}+\tilde{n}_i \pm \sqrt{2}M_i)\notag\\ 
&\hspace{-1 cm}+  f(\tilde{n}_i - \tilde{T}) + h_\pm(M_e \pm \sqrt{\mu}\tilde{n}_e)\,,
\label{Casimir}
\end{align}
where $g_\pm, f$ and $h_\pm$ are arbitrary functions.  

\section{NORMAL FIELDS}\label{sec:normal fields}

The general form of the Casimir \eqref{Casimir} suggests the introduction of a new set of variables which are called  ``normal fields" (see e.g.\ Refs.\ \onlinecite{thiffeault2000classification,tassi2008hamiltonian,tassi_MGP2010}):
\begin{align}
\mathcal{V}_{i,\pm} =& \tilde{T}+\tilde{n}_i \pm \sqrt{2}M_i\\
\mathcal{V}_{i,f} =& \tilde{n}_i - \tilde{T}\\
\mathcal{V}_{e,\pm} =& M_e \pm \sqrt{\mu}\tilde{n}_e\,. 
\end{align}
We claim that if we  express the equations of motion \eqref{n_i} - \eqref{M_e} and the bracket of  \eqref{bracket} in terms of these fields, they will take a simple form. To do so, the following chain rule expressons for  functional derivatives in terms of these new fields are required:
\begin{align}
F_{n_i} =& F_{\mathcal{V}_{i,+}} + F_{\mathcal{V}_{i,f}} + F_{\mathcal{V}_{i,-}}\label{chain1}\\
F_{T_\parallel} =& F_{\mathcal{V}_{i,+}}+F_{\mathcal{V}_{i,-}}-F_{\mathcal{V}_{i,f}}\\
F_{M_i}=& \sqrt{2}\left(F_{\mathcal{V}_{i,+}}-F_{\mathcal{V}_{i,-}}\right)\\
F_{M_e} =& F_{\mathcal{V}_{e,+}}+F_{\mathcal{V}_{e,-}}\\
F_{n_e} =& \sqrt{\mu} \left(F_{\mathcal{V}_{e,+}}-F_{\mathcal{V}_{e,-}}\right)
\label{chain5}\,. 
\end{align}
Using  \eqref{chain1}-\eqref{chain5} the Poisson bracket of  \eqref{bracket} becomes 
\begin{align}
\{F,G\} = &-2 \big{<}\mathcal{V}_{i,f}[F_{\mathcal{V}_{i,f}},G_{\mathcal{V}_{i,f}}]\notag\\ 
&\hspace{-.4 cm}+2\left(\mathcal{V}_{i,+}[F_{\mathcal{V}_{i,+}},G_{\mathcal{V}_{i,+}}] + \mathcal{V}_{i,-}[F_{\mathcal{V}_{i,-}},G_{\mathcal{V}_{i,-}}]\right)\notag\\
&\hspace{-.4 cm} -\sqrt{\mu}\left(\mathcal{V}_{e,+}[F_{\mathcal{V}_{e,+}},G_{\mathcal{V}_{e,+}}] - \mathcal{V}_{e,-}[F_{\mathcal{V}_{e,-}},G_{\mathcal{V}_{e,-}}]\right)\notag\\
&\hspace{-.4 cm} +2\sqrt{2}\left(F_{\mathcal{V}_{i,+}}\partial_z G_{\mathcal{V}_{i,+}} - F_{\mathcal{V}_{i,-}}\partial_z G_{\mathcal{V}_{i,-}}\right)\notag\\
&\hspace{-.4 cm}-\sqrt{\mu}\left(F_{\mathcal{V}_{e,+}}\partial_z G_{\mathcal{V}_{e,+}} - F_{\mathcal{V}_{e,-}}\partial_z G_{\mathcal{V}_{e,-}}\right)\big{>}\,.
\end{align}
This simple form of the bracket is called a direct product\cite{thiffeault2000classification}, and its form  immediately ensures the Jacobi identity.  Since the  inner brackets satisfy the Jacobi identity, so do their sums which constitute the larger bracket of Eq.\eqref{bracket}.

Having expressed the bracket in terms of the normal fields, we can now write down the equations of motion that these fields satisfy, viz. 
\begin{align}
&\frac{\partial \mathcal{V}_{i,\pm}}{\partial t} + [\mathcal{A}_{i,\pm},\mathcal{V}_{i,\pm}] \pm \sqrt{2}\partial_z\mathcal{A}_{i,\pm}= 0\\
&\frac{\partial \mathcal{V}_{e,\pm}}{\partial t} + [\mathcal{A}_{e,\pm},\mathcal{V}_{e,\pm}] \mp \sqrt{\mu}\partial_z\mathcal{A}_{e,\pm}= 0\\
&\frac{\partial \mathcal{V}_{i,f}}{\partial t} + [\mathcal{A}_{i,f},\mathcal{V}_{i,f}] = 0\,,
\end{align}
where
\begin{align}
\mathcal{A}_{i,\pm} =& \Phi + n_i + T \pm \sqrt{2}u_i\\
\mathcal{A}_{i,f} =& \Phi + n_i - T\\
\mathcal{A}_{e,\pm} =&\pm\left(\frac{n_e}{\tau}-\phi\right) + \mu^{\frac{3}{2}}u_e
\end{align}
are stream-functions that simply convect the fields $\mathcal{V}_{s,\pm/f}$. The latter are therefore Lagrangian conserved quantities. Note that in a turbulent system, equipartition results in the flattening of the profiles of Lagrangian invariants.\cite{naulin1998equipartition}

\section{LINEAR STUDY}\label{sec:linear study}

In this section we linearize  \eqref{n_i}-\eqref{M_e} and  the two closure relations \eqref{Ampere}-\eqref{quasineutrality} about an inhomogeneous equilibrium configuration.  Then,  after deriving the dispersion relation, we study the linear stability of the ITG mode. We assume that the densities and temperature vary linearly in the $x$ direction, i.e.,  that these quantities  have  the form $f = x/L_f + \delta f$  with $\delta f = \hat{f} \text{exp}(i {\bf k}\cdot{\bf x} - i\omega t)$. This may be interpreted as a local study, in the WKB sense, for modes satisfying $k_\bot L_\bot\gg 1$ and $k_\| L_\|\gg 1$, where $L_\bot$ and $L_\|$ represent equilibrium scale-lengths.  Our  purpose is to obtain some physical understanding of our model and see how accurately it can describe the various modes of interest.   Next, we assume $\Phi_{eq} = 0$ and $\nabla \psi_{eq} \times \hat{z}= B_{oy} \hat{y}$ with $B_{oy} = -\frac{\partial\psi}{\partial x}$ a constant and $u_{i,eq} = 0$. We note that in Fourier space, the operator $\Gamma_o$ is $\Gamma_o(b) = e^{-b}I_o(b)$, where $b = k^{2}_{\perp}\rho^{2}_i$ (or $b\equiv k^{2}_\perp$ in our normalized units). Even though we mentioned that the model is Hamiltonian only for the choice $\gamma=2$, in the following we keep $\gamma$ general to investigate its effect on the behavior of the modes and we subsequently set $\gamma=2$, to recover the results for  our model.

Moreover, we add two dissipative terms to Eq.~\eqref{T} that are related to the parallel and toroidal resonances. Therefore, from now on, we make the distinction between the non-dissipative, i.e.\ Hamiltonian, gyrofluid model and the one where dissipation terms are included. 

Parameters $\chi$ and $\nu$ of the added dissipative terms are tuned so that the response function of a gyrofluid model matches the kinetic one in the slab and the toroidal limits,  respectively. Their values have been computed in Refs.~\onlinecite{hammett1990fluid,waltz1992gyro} and found to be $\chi = \frac{2}{\sqrt{\pi}}$ and $\nu = 2.019$. Although the $\chi$ value is exact, the numerical value of $\nu$ has not been calculated for the particular model we are presenting but for a similar gyrofluid model. Nevertheless, we will adopt it. The reason is that here, we are mainly concerned with the non-dissipative, Hamiltonian part of the model and the addition of the dissipative terms is not intended to enhance the accuracy of the results,  but merely to show the reader that such a modification is indeed possible. Correct treatment of dissipation would require the proper study of the response function of a kinetic model containing the same physics and the numerical minimization of the error in matching it with the response function obtained by  \eqref{n_i}-\eqref{M_e}.  Such a study is beyond the goals of this paper.

The linearization of the equations of motion and the closure relations in Fourier space result in the following system of equations: 
\begin{align}
-\omega\hat{n}_i =&  \omega_{\ast} \Gamma^{1/2}_0(b)\hat{\phi} -  k_y  B_{oy} \hat{u}_i\notag\\
& -2 \omega_{\ast}\epsilon(\hat{n}_i + \Gamma^{1/2}_0(b)\hat{\phi}+\hat{T})\notag\\
&-  k_z \hat{u}_i \,,
\label{n_i-hat}
\end{align} 
\begin{align}
-\omega(\Gamma^{1/2}_0(b)\hat{\psi} + \hat{u}_i) =& -  k_y B_{oy} \hat{T} -  \omega_{\ast}\eta_i \Gamma^{1/2}_0(b)\hat{\psi}\notag\\
& -  k_y B_{oy} \Gamma^{1/2}_0(b)\hat{\phi}  -  k_y B_{oy} \hat{n}_i \notag\\
&-  \omega_{\ast} \Gamma^{1/2}_0(b)\hat{\psi} -4 \omega_{\ast}\epsilon\hat{u}_i\notag\\
&-  k_z  \hat{n}_i -  k_z \hat{T} \notag\\
&-  k_z \Gamma^{1/2}_0(b)\hat{\phi}\,,
\label{M_i-hat}\\
-\omega\hat{T} =&  \omega_{\ast}\eta_i \Gamma^{1/2}_0(b)\hat{\phi}  - (\gamma-1) k_y B_{oy} \hat{u}_i \notag\\
&-2 \omega_{\ast}\epsilon (\hat{n}_i + \Gamma^{1/2}_0(b)\hat{\phi}+\hat{T})\notag\\
&-  (\gamma-1)k_z \hat{u}_i +2i\nu |\omega_{\ast}|\epsilon \hat{T} 
\label{T_i-hat}\\
&+i\chi|k_{\parallel}|\hat{T}\,,
\notag\\
-\omega \hat{n}_e =&  \omega_{\ast} r_n \hat{\phi} -  k_y B_{oy} \hat{u}_e \notag\\
&+2 \omega_{\ast}\epsilon\left(\frac{\hat{n}_e}{\tau} - \hat{\phi}\right)-  k_z \hat{u}_e\,,
\label{n_e-hat}\\
-\omega(\hat{\psi} - \mu\hat{u}_e) =& - k_y B_{oy} \hat{\phi}  +  \frac{\omega_{\ast}r_n}{\tau}  \hat{\psi}  
+  \frac{k_y}{\tau} B_{oy} \hat{n}_e 
\notag\\
&+2 \omega_{\ast}\epsilon\mu \hat{u}_e -  k_z \hat{\phi} +  \frac{k_z}{\tau} \hat{n}_e\,,
\label{M_e-hat}\\
\hat{n}_e =& \Gamma^{1/2}_0(b)\hat{n}_i + (\Gamma_0(b)-1)\hat{\phi}\,,
\label{Ampere-hat}\\
\frac{2}{\tau \beta_e}k^2_{\perp} \hat{\psi} =& -\hat{u}_e + \Gamma^{1/2}_0(b)\hat{u}_i\,.
\label{quasineutrality-hat}  
\end{align} 

Note that  $\Gamma^{1/2}_0(b)B_{oy} = B_{oy}$ and, to be clear,  recall  the ion and electron density and parallel temperature gradients vary linearly, i.e., $n_i = {x}/{L_{n_i}}, n_e =  {x}/{L_{n_e}}$, and  $T = {x}/{L_T}$.   We simplify the result by setting $k_\parallel = k_z + B_{oy}k_y$ and by defining the parameters $\eta_i =  {L_{n_i}}/{L_T}$, $\epsilon = u_d L_{n_i}$, and $r_n =  {L_{n_i}}/{L_{n_e}}$.   Also, $\hat{\omega}_{\ast} = ({c T_e}/{e B_o})({\hat{k}_y}/{L_n})$ is the usual diamagnetic frequency. In dimensionless variables it is expressed as $\omega_{\ast} = \tau u_{ti}   k_y/L_n$.

\subsection{ELECTROSTATIC DISPERSION RELATION}

The electrostatic limit, which is applicable for low-$\beta$ conditions,\cite{horton1983electromagnetic} leads to a cubic dispersion relation that  offers the opportunity of comparing analytic solutions of the gyrofluid model to kinetic results. To make contact with well-known analytic results for the slab branch of the ITG mode, we also neglect toroidal effects. That is, we drop  all toroidal terms of  Eqs.~\eqref{n_i-hat}-\eqref{M_e-hat}, set $\hat{\psi} = 0$,  and study the slab, electrostatic ITG modes, where the drive is due to the coupling of the parallel transit of particles with the temperature gradient. We notice that in this case, the electron and ion fields are decoupled so we only use the ion field of Eqs.~\eqref{n_i-hat}-\eqref{T_i-hat},  along with the quasineutrality condition of  \eqref{quasineutrality-hat} and the electron adiabatic response $n_e \approx  {\phi}/{\tau}$.  After straightforward manipulations, we obtain a dispersion relation with real part given by 
\begin{align}
&\left(\frac{1}{\tau}+1-\Gamma_o(b)\right)\omega^3  
+ \gamma k^2_\parallel\left(\Gamma_o(b)\frac{\gamma-1}{\gamma}-\frac{1}{\tau}-1\right) \omega  \notag\\
&\hspace{.5 cm} - \Gamma_o(b)\omega_{\ast}\omega^2 + \Gamma_o(b) k^2_\parallel \omega_{\ast}\left((\gamma-1)-\eta_i\right)=0\label{electrostatic_disp_rel}
\end{align}
and  imaginary part by 
\begin{align}
 &k_\parallel\left(\frac{1+(1-\Gamma_o(b))\tau}{\tau}\omega^2 +\Gamma_o(b)\omega_{\ast}\omega
 \right. \notag\\
 &\hspace{4 cm} \left. - \frac{k^2_\parallel (1+\tau)}{\tau}\right)=0\,.
  \label{imaginary}
\end{align}

A simple picture of the dynamics of the ITG instability, as determined by e.g.\ \eqref{electrostatic_disp_rel},  is given in \rcite{cowley1991considerations}, a picture that will be helpful for interpreting  our results. A basic scenario for the development of the instability can start with a density perturbation, which is confined to  variation along the field line because the $E\times B$ velocity across the field lines is incompressible. Electrons respond adiabatically to this  ion density perturbation in order to maintain quasineutrality and in doing so set up an electrostatic potential. This potential perturbation then leads to an $E\times B$ drift that injects  cool ions into the compressed (increased density) region, thereby {\em lowering} the pressure. That is, the plasma exhibits negative compressibility. The resulting lowered pressure then draws ions along the field lines by generating a $u_\parallel$. Since the whole picture develops  in time and moves perpendicular to both the magnetic field and the temperature gradient, the ions that move parallel to the field line,  prompted by the lowered pressure, end up increasing the initial density perturbation.

\begin{figure}[htbp]
\includegraphics[scale=0.9]{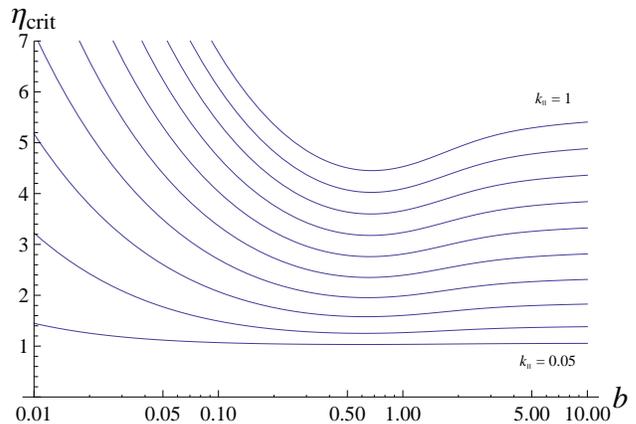}
\vspace{0.5cm}
\caption{Stability criterion with finite $k_\perp$ as given by $b$. Here $k_\parallel$ ranges from 0.05 to 1.0}
\label{kin_lim}
\end{figure}

Returning to Eq.~\eqref{electrostatic_disp_rel} we can infer two stability criteria. 
The first one comes from neglecting the dissipative terms, hence having just the real part of the dispersion relation and by demanding the third-order polynomial to have only real roots. This is done by setting the cubic discriminant equal to zero and by that deriving a quadratic equation in $\eta_i$. To investigate the case of finite $k_\perp$, we obtain the stability criterion by making no approximation on $\Gamma_o(b)$. The result is shown in Fig.~\ref{kin_lim} where $\eta_{crit}$ (the root of the quadratic equation mentioned above) has been plotted as a function of $b$ for various values of $k_\parallel$. 
The curves depicted in  Fig.~\ref{kin_lim} are qualitatively similar to  those  reported in \rcite{antonsen1979inward} where a kinetic model was used. 

\begin{figure}[htbp]
\includegraphics[width=.4\textwidth]{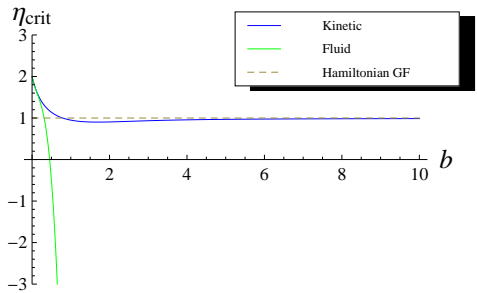}
\vspace{0.5cm}
\caption{Comparison of critical $\eta$ between kinetic, fluid and gyrofluid results for the case $k_\parallel = 0$}\label{eta_crit}
\end{figure}

The second stability criterion we deduce, concerns the case of perturbations with very long parallel wavelengths and comes from setting the imaginary part of the dispersion relation equal to zero, solving for $\omega$ under the condition $k_\parallel = 0$,  and eliminating it from Eq.~\eqref{electrostatic_disp_rel}.  With this procedure we find  
\begin{equation}
\eta^{GF}_{crit} = \gamma-1\,.
\end{equation}
Observe,  the critical value depends on the adiabatic index. The kinetic result for this limiting case is provided in \rcite{kadomtsev1995turbulence} and is given by 
\begin{equation}
\eta^{KIN}_{crit} = \frac{2}{1+2b\left(1-\frac{I_1(b)}{I_o(b)}\right)}\,, 
\end{equation}
with $I_1(b)$ and $I_0(b)$ being modified Bessel functions of the first kind. Note that the adiabatic index in the exact moment equation for the evolution of the parallel temperature is 3.
In Fig.~\ref{eta_crit} we plot this relation and the corresponding fluid approximation of it and we notice that our gyrofluid model has the correct asymptotic behavior for perturbations with very small perpendicular wavelengths provided $\gamma=2$. However, had we chosen  $\gamma=3$, we would have gotten the correct asymptotic behavior for very large perpendicular wavelengths, at the cost of a non-Hamiltonian model. Moreover, the choice $\gamma =  {5}/{3}$ gives $\eta_{crit} = {2}/{3}$, the  result for the fluid model of \rcite{antonsen1979inward}. 

The reason behind this discrepancy stems from the fact that our model lacks an equation for the evolution of the perpendicular temperature. Therefore, all assumptions about the correlation of $T_\perp$ and $T_\parallel$ are made by the choice of $\gamma$ (with $\gamma = 3$ meaning $T_\perp$ and $T_\parallel$ are uncorrelated and $\gamma = {5}/{3}$ meaning $T_\perp = T_\parallel$) and remain fixed throughout the dynamics. 
Despite this obvious inflexibility of the gyrofluid model, it is evident from Fig.~\ref{eta_crit} that it still remains far superior compared to its FLR counterpart.  

\begin{figure}[htbp]
\includegraphics[width=.4\textwidth]{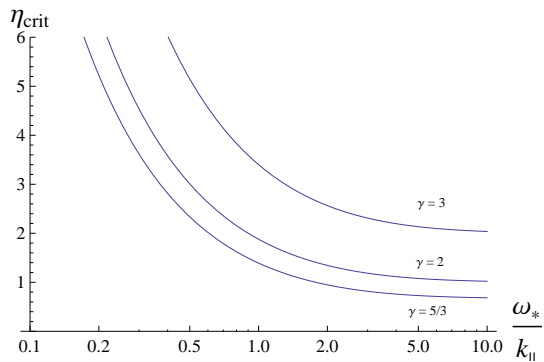}
\vspace{0.5cm}
\caption{Stability criterion at the `fluid' limit with $\tau = 1$ for different values of the adiabatic index}\label{fluid_lim}
\end{figure}

It is helpful to study the  `fluid' limit  of Eq.~\eqref{electrostatic_disp_rel}, which is   obtained by setting $\Gamma_o(b) = 1$ corresponding to  $b = 0$.  This is  the limit of very long perpendicular wavelengths compared to the gyroradius.  Figure \ref{fluid_lim} shows the stability criterion in this fluid limit for three different values of $\gamma$, results  that were previously  obtained in \rcite{antonsen1979inward} for $\gamma = {5}/{3}$, where  a   heuristic explanation was  for  given for the  $\eta_{crit}$ limiting value for very long $k_\parallel$. 

To conclude with the electrostatic slab case, we investigated the growth rate as a function of $\tau$. The condition  $\tau < 1$ or,  in other words, $T_i > T_e$ is a well-known  stabilizing factor for ITG,  which is of particular importance for the hot-ion cores of tokamaks.\cite{hahm1988properties,romanelli1989ion,dong1992toroidal} Indeed, the behavior we found was the expected one.

\subsection{ELECTROMAGNETIC DISPERSION RELATION}

To be applicable to the higher plasma pressure achieved by auxiliary or alpha-particle heating, the theory must include the electromagnetic effect.  In fact, this effect becomes important at surprisingly low-$\beta$ because of other small parameters in the problem. It is well known that increasing $\beta$ stabilizes ITG modes\cite{kim1993electromagnetic},  but leads to the onset of kinetic ballooning modes, also known as the Alfv\'{e}nic ITG modes (AITG).\cite{ZoncAITG99} For toroidal ITG modes, the drive comes from the coupling of curvature and $\nabla B$-drift terms with the temperature gradient, so that we must now keep the toroidal curvature terms. It can be easily seen that, to lowest order, the electromagnetic effect is stabilizing. The electromagnetic perturbation creates a small component of \textbf{B} that is perpendicular to both the background magnetic field and the pressure gradient. This component then leads to the development of a force on the ions, parallel to the field lines that opposes the attraction from the pressure lowering of ITG.


In the remainder of this section, we follow the analysis of Kim, Horton and Dong\cite{kim1993electromagnetic} and compare our gyrofluid results with their local kinetic ones. We note, however, that complete agreement cannot be expected since  Kim {\it et al.}\  has one extra parameter, namely $\eta_e$. We also note that the eigenfrequencies for the model in \rcite{snyder2001landau} lie within a few percent of the kinetic results, so that comparing our model to the kinetic results is effectively equivalent to comparing it to the Snyder and Hammett model.

\begin{figure}[htbp]
\centering
\begin{subfigure}[b]{0.5\textwidth}
\includegraphics[scale=0.9]{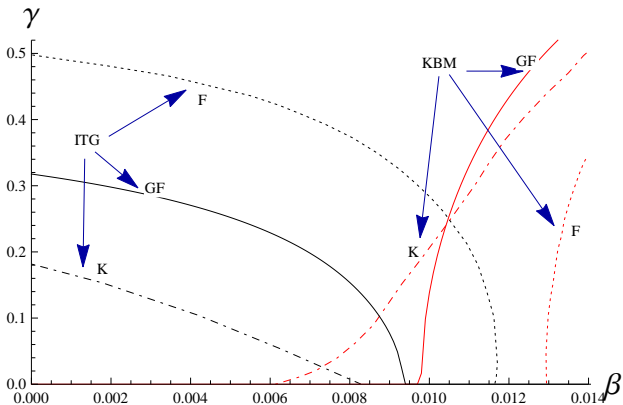}
\vspace{0.5cm}
\caption{Hamiltonian model without dissipative terms and comparison with kinetic and fluid results.}\label{em_growth_without}
\end{subfigure}
\begin{subfigure}[htbp]{0.5\textwidth}
\includegraphics[scale=0.9]{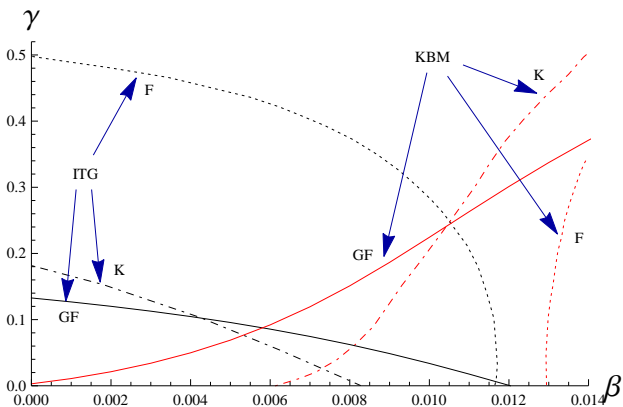}
\vspace{0.5cm}
\caption{Model with dissipative terms and comparison with kinetic and fluid results.}\label{em_growth_with}
\end{subfigure}
\caption{Normalized growth rate vs. $\beta$ for $\eta_i = 2.5, b = 0.5, k_\parallel =0.1, \epsilon = 0.2, r_n = 1$, and $\tau=1 $. Toroidal ITG is the  black line while  KBM is the red.}\label{em_growth}
\end{figure}

Because the dispersion relation becomes unwieldy and doesn't provide much physical insight, we refrain from displaying it here. Instead, we solve it numerically and present the results.   Figure \ref{em_growth} shows the normalized growth rates for (a) the ideal and (b) the ``Landau'' versions of our model as a function of $\beta$ when $\eta_i = 2.5, b = 0.5, k_\parallel = 0.1, \epsilon = 0.2, r_n = 1$ and  $\tau=1$. We also provide the kinetic and fluid model results from \rcite{kim1993electromagnetic} for comparison.  By the  ``Landau'' version we mean of course the Hamiltonian model augmented by dissipative terms modeling the damping caused by the wave-particle interactions. From Fig.~\ref{em_growth_without}, it becomes immediately clear that the nonlocal treatment of the ion response in the gyrofluid model reproduces the main qualitative features of the kinetic result much better than the fluid model. Compared to the fluid result, the gyrofluid one gives stronger stabilization of the ITG modes and lower thresholds for the excitation of KBMs. This is related to the toroidal resonance.
In both Fig.~\ref{em_growth_without} and Fig.~\ref{em_growth_with} we observe the close connection between the stabilization of the ITG mode and the excitation of the Kinetic Ballooning mode in accordance with what kinetic theory predicts.  The addition of dissipative terms makes the curves shift closer to the kinetic result although we remark that at low growth rates the agreement is less satisfactory.  

\begin{figure}[htbp]
\includegraphics[scale=0.9]{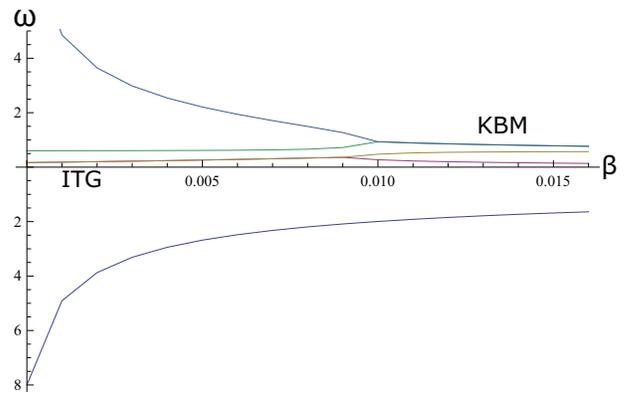}
\caption{Real frequency vs. $\beta$. All parameters are the same as in Fig.4}
\label{real_f}
\end{figure}

Here, we pause to explain an interesting effect,    the destabilization  due to the addition of dissipation of  two previously marginally stable modes ($\gamma=0$).   For example, for the GF model it is seen in  Fig.~\ref{em_growth_without}   that without dissipation  the KBM becomes unstable at   $\beta\approx0.010$, while in   Fig.~\ref{em_growth_with} it is seen  for the  same case with dissipation that this mode is destabilized for all values of $\beta$.  A similar shift from stability to instability can be observed upon comparing  these figures for the ITG mode, which  is seen in Fig.~\ref{em_growth_without}  to  transition to instability at a somewhat smaller value of  $\beta$.  To understand these transitions consider  Fig.~\ref{real_f}, where we plot the real parts of the frequencies  versus $\beta$ for four modes of the GF model  without dissipation.  In this figure the two upper most modes are marginally stable for small values of $\beta$, then as $\beta$ approaches  the transition value near $0.010$ they collide and produce instability.   This   bifurcation, which is standard in Hamiltonian systems, is  called the  Hamiltonian Hopf (or Kre$\breve{\i}$n) bifurcation\cite{morrison1998hamiltonian,Hagstrom14}.   Observe, the same bifurcation occurs when two marginally stable modes  collide as  $\beta$ is decreased to a value near to the KBM transition but closer to $0.009$, producing  the unstable ITG mode.  (After the transitions there are also damped modes that are not shown in the figures.)

The dissipative destabilization observed in  Fig.~\ref{em_growth_with}  is a generic feature of Hamiltonians systems with negative energy modes (NEMs). Indeed, in Hamiltonian systems, whenever a Hamiltonian Hopf bifurcation occurs, one of the modes {\it must} be a NEM, and such modes have the property of getting destabilized with the addition  of dissipation (see e.g.\ Ref.~\onlinecite{kotsch90} for a Hamiltonian version  of the classical Kelvin-Tait theorem\cite{Tait21}).   One could perform a calculation like those of 
Refs.~\onlinecite{tassi2008hamiltonian,tassi2011mode}, where the modal eigenvector is inserted into the perturbation energy in order to show explicitly that it is an NEM and then show the dissipative terms remove energy from this mode, but such a calculation is outside the scope of the present paper.   (A similar situation happens when energy is added to a positive energy mode.)  Also note, the previously unstable ITG and KBM modes of Fig.~\ref{em_growth_without} become less unstable at the onset of dissipation, as is shown in Fig.~\ref{em_growth_with}   due to the fact that it becomes harder for a mode to grow when there is less available energy in the system, which is consistent with this scenario.

We reiterate that the purpose of our model is to improve the {\em nonlinear} fidelity of fluid models. From that perspective, we view the quality of agreement in Fig.~\ref{em_growth} as adequate.



\begin{figure}[htbp]
\includegraphics[scale=0.9]{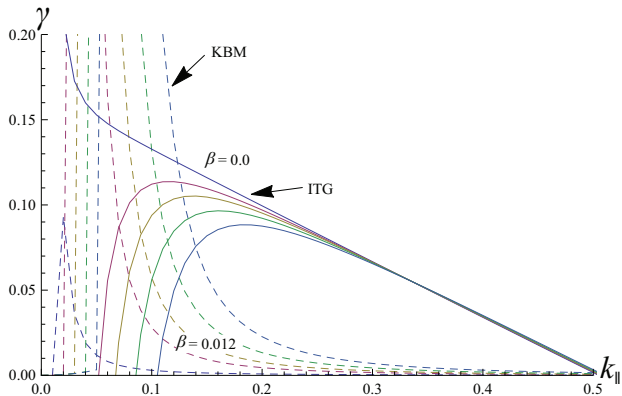}
\captionof{figure}{Growth rates of the ITG-KBM modes as a function of the parallel wavenumber for the gyrofluid model with dissipative terms.}\label{em_kpar_with}
\end{figure}

In Fig.~\ref{em_kpar_with} we display the dependence of the growth rates of the ITG and KBM modes on $k_\parallel$ for various values of $\beta$ for the model augmented with the dissipative terms, with all other parameters remaining the same as in Fig.~\ref{em_growth}. Values of $\beta$ are in the range  $0.000-0.012$. We observe that the stabilization through the electromagnetic effect becomes more efficient with decreasing $k_\parallel$. Further, we see again the near simultaneous stabilization of ITG and destabilization of KBM as was noted above. 

For large values of $k_\parallel$ the mode is stabilized by the large parallel ion transit term.\cite{dong1992toroidal,kim1991transition,dominguez1989local} Intuitively, we can understand that the mode is limited by the fact that an appreciable initial density perturbation cannot be created within an arbitrarily small length scale. Even before this limit is reached,  though, the negative compressibility mentioned above is proportional to the ratio of the ion diamagnetic to the sound frequency ($\omega_{pi}/k_\|c_s$), so that coupling to the sound wave acts as a source of stabilization. On the other hand, the initial density and potential perturbations, as well as the resulting pressure lowering,  are all proportional to $k_\parallel$. Therefore, a finite $k_\parallel$ is needed to overcome the stabilizing effect of curvature and $\beta$. Thus, the mode becomes most unstable at some intermediate value. We remark that in practice a complete treatment of the effect of $k_\parallel$ requires a nonlocal approach since $k_\|L_\|\sim 1$ normally applies, so that the WKB approach in insufficient.

\begin{figure}[htbp]
\includegraphics[scale=0.9]{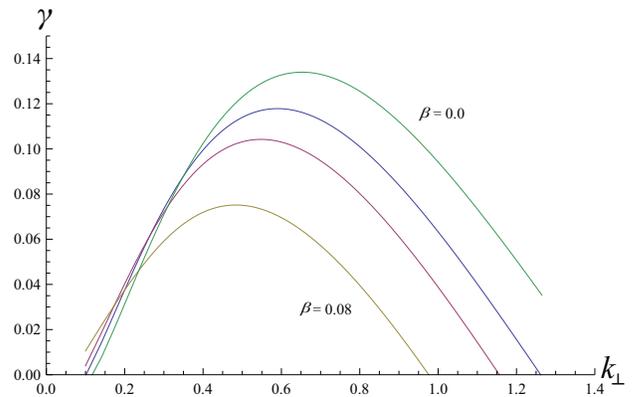}
\captionof{figure}{Growth rates of the ITG-KBM modes as a function of the perpendicular wavenumber for the gyrofluid model with dissipative terms.}\label{em_kperp_with}
\end{figure}

In Fig.~\ref{em_kperp_with} we illustrate the behavior of the growth rate versus $k_\perp$ for various values of $\beta$, with the same parameters as those of the previous figures. We notice that the peak growth rate occurs around  $k_\perp\approx 0.65$ and does not  change much with $\beta$. Furthermore, the stabilizing effect of $\beta$ is almost uniform for wavenumber values higher than this. This could be attributed to a very high phase velocity of the wave,  which leaves few particles with the right thermal speed to resonate with it. For smaller wavenumbers, however, the stabilization due to $\beta$ becomes ineffective. This is because in this region  the parallel ion transit term becomes significantly larger than the curvature term and becomes the dominant stabilizing effect. Another important stabilizing effect at high $k_\perp$ comes from FLR physics. Namely, the ions respond to the gyroaveraged electrostatic field, thereby reducing the effective $E\times B$ velocity.  

\begin{figure}[htbp]
\includegraphics[scale=0.9]{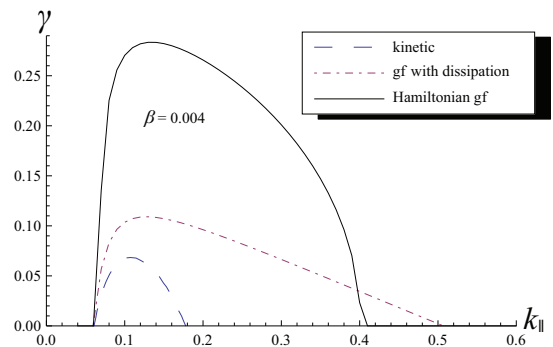}
\captionof{figure}{Comparison between growth rates of the ITG mode as a function of the parallel wavenumber for the ideal Hamiltonian model, the model with linear wave-particle (Landau) damping, and the kinetic results.}\label{em_kpar_comparison}
\end{figure}

\begin{figure}[htbp]
\includegraphics[scale=0.9]{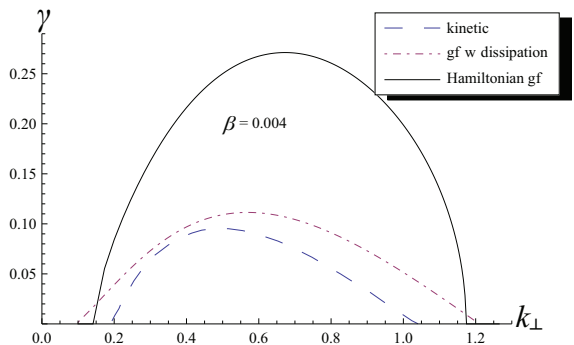}
\captionof{figure}{Comparison between growth rates of the ITG mode as a function of the perpendicular wavenumber for  the ideal Hamiltonian model, the model with linear wave-particle (Landau) damping, and the kinetic results.}\label{em_kperp_comparison}
\end{figure}

In Figs.~\ref{em_kpar_comparison} and  \ref{em_kperp_comparison} we compare the Hamiltonian model augmented by dissipative terms and the kinetic result from \rcite{kim1993electromagnetic}. Both figures suggest some common features: again, the qualitative similarity between the Hamiltonian and the kinetic curves is evident. However, there is a quantitative disparity since the Hamiltonian result is roughly three times higher than the kinetic one at the peak value of $\gamma$. This deviation seems to be corrected by taking into account the dissipative terms which lowers  the results to at most 30\% off from the kinetic ones  at peak growth rate. This amendment, though,  doesn't come without its own problems, namely the erratic behavior of the dissipative model at low values of $\gamma$.

\section{CONCLUSIONS}\label{sec:Conclusions}

We presented a Hamiltonian, five-field, electromagnetic gyrofluid model that  evolves three moments for the ions (density, parallel momentum,  and parallel temperature) and two moments for the electrons. We gave the Hamiltonian formulation of the model by providing a suitable Hamiltonian and a Lie-Poisson bracket that satisfies the Jacobi identity. For this system, we found the families of Casimir invariants and,  from them, we defined five normal fields in terms of which the equations of motion and the bracket take their simplest form.

To evaluate the physical fidelity of the model, we performed a local, linear study of the dispersion relation. We began by comparing the electrostatic dispersion relation with known analytic results for a slab ITG mode. We found that the critical $\eta_i$ for the onset of instability is unity for all $k_\perp$. This is very close to the kinetic result for perpendicular wavelengths comparable to and greater than the ion gyroradius, but it is only half of the exact value (two) for long wavelengths. Ordinary fluid models, by contrast, yield good agreement at long wavelength but predict negative values of $\eta_{i,\mbox{\scriptsize crit}}$ at moderate and short wavelengths. By ordinary fluid models, we refer here to those that are derived from the Braginskii model and other long wavelength expansion procedures.

We subsequently examined the electromagnetic properties of the model, including toroidal curvature, by comparing the dependence of the growth rate on well known stabilizing factors and comparing them with the local kinetic result of \rcite{kim1993electromagnetic}. We found good qualitative agreement although the two models cannot be directly compared since the latter includes the effects of the electron temperature gradient. We leave for future work the task of including in the model an electron temperature evolution equation in a manner that preserves the Hamiltonian character. 

Given the wide availability of several high-quality gyrokinetic (GK) codes that have been verified and validated in a broad array of contexts, it is appropriate to reflect on the value of gyrofluid (GF) models. Due to its nature as a truncated moment expansions of the GK model, a GF model such as the one presented here cannot aspire to compete with the latter in any but three domains: speed, ease of use, and by virtue of the first two, ability to generate physical insight. The success of the TGLF code\cite{TGLF05,Holl13,KinStaeb11} demonstrates that there is a strong demand for an agile quasilinear GF code to understand and interpret experimental observations of turbulent transport. The motivation for the development of the Hamiltonian GF model presented here is similar but different: it is to provide an equally agile tool to investigate {\em multi-scale nonlinear} problems such as those listed in the introduction. In this context, the linear accuracy of the model is of secondary importance compared to assuring the proper conservation laws and providing a qualitatively correct picture of the nonlinear energy transfers. It is worth noting, in this context, that a Poisson bracket for a gyrokinetic model, demonstrating its Hamiltonian nature, has only recently been constructed\cite{burby2014Ham} using the  newly developed technique of gauge-free lifting.\cite{GFreeLift13}

Uncertainty quantification (UQ) provides another application for reduced models that is worth mentioning. The large number of inputs to gyrokinetic codes make comprehensive UQ impractical, but the existence of a reduced model opens up new avenues for charting model sensitivities and subsequently using Bayesian inference with a smaller number of runs of the GK code to selectively refine the predictions and reduce the error bars.


\section*{Acknowledgements}

This material is based upon work supported by the U.S. Department of Energy, Office of Science, Office of Fusion Energy Sciences, under Award Number DE-FG02-04ER-54742. I.K.C would like to thank Ryan White for helpful discussions. One of us (FLW) also wishes to acknowledge support from the Katholieke Universiteit Leuven where the above research was initiated. 

\bibliographystyle{unsrt}

\end{document}